\documentclass[aps,pre,twocolumn,groupedaddress]{revtex4}
\usepackage{amsfonts}
\usepackage{amsmath}
\usepackage{amssymb}
\usepackage{graphicx}%

\begin{document}
\title{Fluctuations of local density of states and $C_0$ speckle correlations are equal}
\author{B.A. van Tiggelen}
\email[]{Bart.Van-Tiggelen@grenoble.cnrs.fr}
\author{S.E. Skipetrov}
\email[]{Sergey.Skipetrov@grenoble.cnrs.fr}
\affiliation{Laboratoire de Physique et Mod\'{e}lisation des
Milieux Condens\'{e}s/CNRS, Maison des Magist\`{e}res/UJF, BP 166,
38042 Grenoble, France.}


\begin{abstract}
We establish a conceptual relation between the fluctuations of the local density of states (LDOS) and intensity correlations in speckle patterns
resulting from multiple scattering of waves in random media. We show that among known types of speckle correlations ($C_1$, $C_2$, $C_3$, and
$C_0$) only $C_0$ contributes to LDOS fluctuations in the infinite medium. We propose to exploit the equivalence of LDOS fluctuations and $C_0$
intensity correlation as a `selection rule' for scattering processes contributing to $C_0$.
\end{abstract}
\maketitle

The local density of states of waves (LDOS) $\rho(\mathbf{r},\omega)$ is an important concept that regularly turns up in discussions of waves in
interaction with media. The number $\rho(\mathbf{r},\omega) dV d\omega $
  represents the local weight  of all eigenfunctions in the frequency interval
$d\omega$ around frequency $\omega$ inside  a small volume $dV$ around position $\mathbf{r}$. In homogeneous media it is independent of
$\mathbf{r}$ and just equal to the `density of states per unit volume' found in all textbooks. Near boundaries the LDOS exhibits Friedel-type
oscillations on the scale of the wavelength \cite{loudon}. In bandgap materials the LDOS was shown to govern the spontaneous emission of an atom
at position $\mathbf{r}$ \cite{sprik,vos1}. In random media, where wave propagation is diffuse, the equipartition principle attributes an
\emph{average} local energy density of radiation that is directly proportional to the ensemble-averaged LDOS. This apparently simple principle
can have surprising consequences, for instance when waves with different velocities participate in the diffusion process, as is the case for
seismic waves \cite{ep}. For disordered bandgap materials \cite{vos} the equipartition principle is surprising in the sense that the multiple
scattering process, with typical length scale the mean free path $\ell$ that is, in general, much larger than the wavelength or the lattice
constant, distributes energy with sub-wavelength structure. From a fundamental point of view, the LDOS is also the crucial quantity in the
recent studies on `passive imaging' \cite{passive}. Its basic principle is that for a homogeneous distribution of sources --- such as noise
--- the field correlation function (with time and space) is essentially proportional to the (Fourier transform of) LDOS,
and thus sensitive to local structure, random or not.

In random media the LDOS is a random quantity. Its statistical distribution has been studied previously within the frameworks of nonlinear sigma
model \cite{wegner80,mirlin}, random matrix theory \cite{beenakker94}, and optimal fluctuation method \cite{smol97}. The purpose of this paper
it to establish a relation between the fluctuations of LDOS --- within the ensemble of random realizations --- and intensity correlations in
speckle patterns. Several contributions to intensity correlations have been identified. The `standard', Gaussian correlation $C_1$ is the best
known \cite{shapiro86}, but non-Gaussian correlations $C_2$ and $C_3$ have been predicted \cite{feng} and observed \cite{speckleobs}, mostly in
the transmitted flux. Recently the $C_0$ has been added \cite{shapiro,sergey}. The $C_0$ correlation is caused by scatterers close to either
receiver or  source and is, surprisingly, of \emph{infinite} spatial range. Contrary to the other correlations, $C_0$ is quite non-universal and
highly dependent on details of the scatterers, such as their phase function. The total transmission coefficient is known to be dominated by
$C_2$, and the conductance by $C_3$. Unfortunately, the basic observable whose fluctuations are dominated by $C_0$ has never been identified.
This is perhaps why its observation has never been reported so far.

\begin{figure}
\includegraphics[width=8.5cm,angle=0]{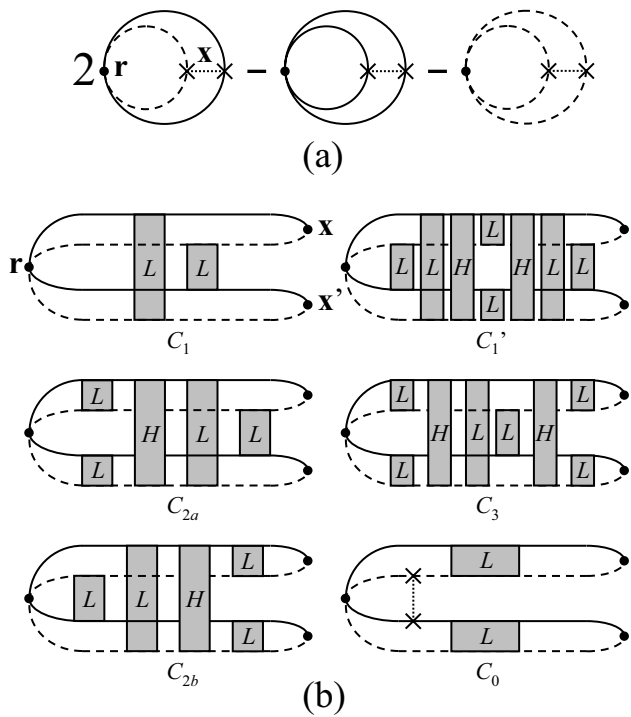}
\caption{\label{ldos} (a). Two-field intensity diagrams that give the leading order to the variance of LDOS in a random medium with Gaussian
uncorrelated disorder. (b). Typical four-field diagrams that contribute to speckle correlations in a random medium. Solid and dashed lines
denote retarded and advanced Green's functions, respectively, shaded boxes are ladder propagators ($L$) and Hikami boxes ($H$), dotted line with
crosses denotes scattering of two wave fields on the same heterogeneity. The variance of the local density of states (LDOS) can be obtained by
integrating over $\mathbf{x}$ and $\mathbf{x}'$. Only the lower right diagram yields a non-vanishing contribution to the variance of LDOS in the
infinite medium.}
\end{figure}

The fluctuations of LDOS can in principle be found from the average (Bethe-Salpeter) two-particle Green's function, but the diffusion approximation that is
usually employed for this object \cite{akkermontam} is not valid on length scales of the order of the wavelength, which appear to give an important contribution. Mirlin \cite{mirlin} noticed that in 3D the result is dominated by nearby scattering  and is of order $1/k\ell$, where $k$ is
the wavenumber and $\ell \gg 1/k$ is the mean free path. An exact calculation in the infinite medium with Gaussian white-noise disorder gives
\begin{eqnarray}\label{ss}
&&\frac{ \mathrm{Var}[\rho(\mathbf{r})]}{\langle\rho(\mathbf{r})\rangle^2} =
\nonumber \\
&&=\left( \frac{4\pi}{k} \right)^2  \frac{1}{2}  \, \left[
   \langle G(\mathbf{r}, \mathbf{r}) G^*(\mathbf{r}, \mathbf{r}) \rangle_c -
\mathrm{Re}\,\langle G^2(\mathbf{r},\mathbf{r})\rangle_c\right] \nonumber \\
&&\approx   \left( \frac{4\pi}{k} \right)^2
 \frac{1}{2} \frac{4\pi}{\ell}\int d\mathbf{x} \, \frac{1-\cos (4kx)}{(4\pi x)^4}=\frac{\pi}{k\ell}
\end{eqnarray}
Here $G(\mathbf{r},\mathbf{r}')$ is the Green's function of the wave equation describing the waves in the random medium. In the second equality
we restricted to single scattering in the Born approximation [see Fig.\ \ref{ldos}(a)]. The value $\pi/k\ell$ agrees \textit{exactly\/} with the
one found for the $C_0$ speckle correlation [lower right diagram in Fig.\ \ref{ldos}(b)] \cite{shapiro}. Going beyond single scattering, i.e.
replacing the dotted lines in Fig.\ \ref{ldos} by diffusion ladders, yields small corrections $\sim 1/(k \ell)^2$ to both the variance of LDOS
and $C_0$. A \emph{deeper, generally valid} relation between the fluctuations of LDOS and $C_0$ correlation is suggested by the above
observations. This is the principal subject of the present work.

Let us consider the simplest model possible: scalar waves in an infinite random medium with white-noise disorder and leave more complicated
situations for future work. Our assumptions are:
\begin{enumerate}
    \item At long distances, the diffusion approximation for the correlation $\langle G(\mathbf{r},\mathbf{x})G^*(\mathbf{r}',\mathbf{x}')\rangle$
    of {two} Green's functions is valid.
    \item The correlation $\langle I(\mathbf{r},\mathbf{x})I(\mathbf{r}',\mathbf{x}')\rangle$  of {two}
    intensities $I(\mathbf{r},\mathbf{x}) \equiv |G(\mathbf{r},\mathbf{x})|^2$ propagating from the source $\mathbf{r}$ to the receiver $\mathbf{x}$
    is composed of  terms of \emph{only} four different
 classes, referred to as $C_1$, $C_2$, $C_3$ and $C_0$, distinguished by a different correlation range.
\end{enumerate}
The first assumption excludes 2D and 1D random media  that are subject to localization effects.  We will thus concentrate on 3D random media.
The classification into $C_1$, $C_2$, $C_3$ and $C_0$ summarizes the outcome of numerous theoretical approaches and experiments
\cite{shapiro86,feng,speckleobs,shapiro,sergey,akkermontam,azi,gabriel}. The class $C_1$ has short range correlation in \emph{both} the source
positions $\mathbf{r}$, $\mathbf{r}'$ and the receiver positions $\mathbf{x}$, $\mathbf{x}'$ (with a range at most equal to the mean free path).
$C_2$ has two parts. The first part has long range correlation (typically a power-law) in source positions and short range correlation in
receiver positions, the second part \emph{vice versa}.  $C_3$ has long range correlation in \emph{both} source and receiver positions. The class
of terms described by $C_0$ exhibits an \emph{infinite } range correlation in \emph{either} source or receiver positions \cite{shapiro,sergey}.
The classes $C_2$, $C_3$ and $C_0$ imply non-Gaussian statistics of the wave field. For weak disorder ($k\ell \gg 1$) this statistics is
Gaussian and $C_1$ dominates.

The random dielectric constant is denoted by $\varepsilon(\mathbf{r})$, and we shall add a fictitious, \emph{{homogeneous}} dissipation
$\varepsilon_a$ and later consider $\varepsilon_a \downarrow 0$. The Green's operator for scalar waves is $G(\mathbf{r},\mathbf{p},\omega)=
[(\varepsilon(\mathbf{r})+i\varepsilon_a)\omega^2/c^2 -\mathbf{p}^2]^{-1}$. The resolvent identity states that $G-G^* =
-2i\varepsilon_a(\omega^2/c^2)\, GG^*$. In real space this translates to the identity
\begin{eqnarray}\label{kirch}
  - \mathrm{Im}\,G(\mathbf{r},\mathbf{r},\omega, \varepsilon_a=0) = \frac{\omega^2}{c^2}\lim_{\varepsilon_a\downarrow 0}
    \varepsilon_a \int d\mathbf{x}\,
    I(\mathbf{r},\mathbf{x})
\end{eqnarray}
where the integral extends over the whole space, and the intensity $I(\mathbf{r},\mathbf{x})$ was defined in assumption 2 above. We recall that
the (radiation) LDOS $\rho(\mathbf{r},\omega)$ is equal to $-(\omega/\pi c^2) \mathrm{Im}\,G(\mathbf{r},\mathbf{r},\omega)$ \cite{sprik}. Thus,
Eq.~(\ref{kirch}) expresses physically that for a homogeneous distribution of sources, the local radiation density is directly proportional to
the LDOS. For brevity we shall drop the frequency reference. The second moment of the LDOS can be expressed as
\begin{eqnarray}\label{var}
\langle \rho(\mathbf{r})^2\rangle &=&
\frac{\omega^6}{\pi^2 c^8}
    \lim_{\varepsilon_a\downarrow 0}
    \varepsilon_a^2 \int d\mathbf{x}\int d\mathbf{x}'\,
\langle I(\mathbf{r},\mathbf{x})
    I(\mathbf{r},\mathbf{x}') \rangle \hspace*{5mm}
\end{eqnarray}

Equation (\ref{var}) establishes a conceptual relation between the variance of LDOS at $\mathbf{r}$, $\mathrm{Var}[\rho(\mathbf{r})] = \langle
\rho(\mathbf{r})^2 \rangle - \langle \rho(\mathbf{r}) \rangle^2$, given by its l.h.s., and intensity correlations in a speckle pattern created
by a point source at $\mathbf{r}$ (the integrand of the r.h.s.). This facilitates a direct correspondence  between the various contributions to
LDOS variance and speckle correlations. We demonstrate in Appendix \ref{appA} that, among the four classes of speckle correlations, only $C_0$
contributes to the LDOS variance, because of it \emph{infinite} range in the receiver positions $\mathbf{x}$ and $\mathbf{x}'$. The contribution
of $C_1$ to Eq.~(\ref{var}) vanishes because $C_1$ is short-ranged. The contributions of the long range correlations $C_2$ and $C_3$ vanish
because they originate from crossings between diffuse propagators that respect current conservation \cite{kane88}.

We conclude that the normalized fluctuations of LDOS and the $C_0$ speckle correlation are one and the same thing:
\begin{eqnarray}
\frac{\mathrm{Var}[\rho(\mathbf{r})]}{\langle\rho(\mathbf{r})\rangle^2} =
C_0
\label{main}
\end{eqnarray}
and that observational attempts to confirm the existence of $C_0$ should focus on the LDOS, either probed by spontaneous emission \cite{sprik}
or by using evanescent waves \cite{greffet}. It follows from our analysis that only correlations with infinite spatial range contribute to
$\mathrm{Var}[\rho(\mathbf{r})]$, and Eq.~(\ref{main}) might serve as a definition for $C_0$. Alternatively, any nonzero variance of LDOS
implies the existence of spatial correlations of the intensity $I$ with infinite range.

Since $C_0$ correlation is  non-universal and sensitive to the local, microscopic structure of the random medium, our Eq.~(\ref{main}) implies
that the fluctuations of LDOS are non-universal too, contrary to the universality of conductance fluctuations. In the context of `imaging with
noise'  \cite{passive}, essentially relying on the  measurement of LDOS, the equivalence of $C_0$ correlation and LDOS fluctuations implies that
only objects closer than a wavelength can affect the LDOS and can thus be imaged.

The $C_0$ correlation determines the variance of LDOS at a given frequency $\omega$, and it continues to do so in the correlation of LDOS at two
frequencies differing by some $\Omega \ll \omega$. We obtain $\langle \rho(\mathbf{r}, \omega) \rho(\mathbf{r}, \omega + \Omega)\rangle_c/
\langle \rho(\mathbf{r}, \omega) \rangle^2  \simeq  C_0$, independent of $\Omega$.  Similarly, if the disordered medium is not stationary, like,
e.g., a suspension of small particles in Brownian motion, the LDOS will fluctuate in time. The time correlation of these fluctuations, $\langle
\rho(\mathbf{r}, t) \rho(\mathbf{r}, t + \tau)\rangle_c/ \langle \rho(\mathbf{r}, t) \rangle^2$, is again determined by $C_0$. According to
Ref.\ \cite{sergey}, $C_0(\tau)$ decays as $\tau^{-3/2}$ for large enough $\tau$. We conclude therefore that LDOS exhibits long-range
correlations in
 time and infinite range  correlations in frequency.

To summarize, our main conclusion is that fluctuations of local density of states for waves in random media are conceptually equal to the
recently predicted, though not yet observed $C_0$ intensity correlation, and not to the other known types of intensity correlation, which have
all been observed. Crucial for this equivalence is the infinite spatial range of $C_0$. In a finite medium the intensity correlations
$C_{1,2,3}$ will emerge as \emph{extensive} contributions to the LDOS variance, vanishing in some way as the medium scales upwards. Our analysis
does not apply to localized media, where all correlation classes might contribute, integrated over the finite volume $\xi^3$, with $\xi$ the
localization length. With some minor modifications our main conclusion should hold for infinite 3D disordered bandgap materials, where the LDOS
is a much less trivial quantity. The equivalence between LDOS fluctuations and $C_0$ intensity correlations can serve as a selection rule for
identifying scattering processes contributing to $C_0$.

This work was supported by GDR 2253 IMCODE of CNRS, NSF/CNRS contract 14872
and ACI 2066 of the Ministry of Research. We thank Michel Campillo,
Roger Maynard and Richard Weaver for enlightening discussions.

\appendix
\section{}
\label{appA}

In this Appendix we demonstrate that $C_1$, $C_2$, and $C_3$ correlation functions do not contribute to the fluctuations of LDOS in Eq.\
(\ref{var}), and that $C_0$ gives the only nonvanishing contribution. We restrict ourselves to infinite, reciprocal media where
$G(\mathbf{x},\mathbf{r}) = G(\mathbf{r},\mathbf{x})$ and assume $k \ell \gg 1$.

We first consider Gaussian ($C_1$) statistics according to which $\langle G(1)G^*(2)G^*(3)G(4)\rangle =\langle G(1)G^*(2) \rangle\langle
G^*(3)G(4)\rangle + \langle G(1)G^*(3)\rangle\langle G^*(2)G(4)\rangle$. The first term just gives the average LDOS squared. In the diffusion
approximation (assumption 1), the correlation of two Green's functions takes the form \cite{pr}:
\begin{eqnarray}\label{twopar}
&&\langle G(\mathbf{r},\mathbf{x})G^*(\mathbf{r}',\mathbf{x}')\rangle =
\nonumber \\
&&\hspace*{5mm}=
\langle -\mathrm{Im}\,G(\mathbf{r},\mathbf{r}')\rangle \, L(\mathbf{r},\mathbf{x}) \, \langle-\mathrm{Im}\,
    G(\mathbf{x},\mathbf{x}')\rangle \hspace*{5mm}
\end{eqnarray}
In the infinite medium, the field propagator $\langle \mathrm{Im}\,G(\mathbf{r},\mathbf{r}')\rangle $ oscillates algebraically on the scale of
the wavelength and decays exponentially beyond the extinction length $\ell$. The ladder propagator $L(\mathbf{r},\mathbf{x})$, however, is very
long-range and decays significantly only by absorption. Therefore, for the purpose of this paper we do not have to discriminate between
$\mathbf{r}$ and $\mathbf{r}'$ or $\mathbf{x}$ and $\mathbf{x}'$ in $L$. On long length scales $L$ obeys a diffusion equation with absorption
time $\tau_a$:
\begin{equation}\label{diff}
    -D \nabla^2 L(\mathbf{r},\mathbf{x})+ \frac{1}{\tau_a}L(\mathbf{r},\mathbf{x}) =
    K \delta(\mathbf{r}-\mathbf{x})
\end{equation}
where the factor $K= \lim_{\varepsilon_a\downarrow 0} \left[
 \pi  \langle \rho(\mathbf{x}) \rangle \omega \varepsilon_a
\tau_a\right]^{-1}$ is imposed by the  ensemble-average of
 Eq.~(\ref{kirch})
\footnote{In statistically homogeneous media $\langle \rho(\mathbf{x}) \rangle
$ is independent of $\mathbf{x}$. In bandgap materials --- in principle
beyond the scope of this work --- the average over both disorder and
unit cell should appear here, and is thus still independent of
$\mathbf{x}$.}.

The variance of LDOS caused by $C_1$ becomes [see the upper left diagram in Fig.\ \ref{ldos}(b)]
\begin{eqnarray}\label{Var1}
    \mathrm{Var}_1[\rho(\mathbf{r})] &=& \frac{\omega^4}{c^4} \lim
    _{\varepsilon_a\downarrow 0}  \varepsilon_a^2  \langle
    \rho(\mathbf{r})\rangle^2 \int d\mathbf{x} \, L^2(\mathbf{r},\mathbf{x})
\nonumber \\
&\times& \int d\Delta \mathbf{x}\,  \langle -\mathrm{Im }\, G(\Delta \mathbf{x}) \rangle ^2
\end{eqnarray}
The integrand of the second integral is $\sin^2(k\Delta x)\exp(-\Delta x/\ell)/(\Delta x)^{2}$, making the integral converge after typically the
extinction length $\ell$, without the need for absorption. The integrand of the first integral is typically $L(x)^2 \sim (K^2/D^2x^2)\exp
(-2x/\sqrt{D\tau_a})$. The critical contribution of Eq.~(\ref{Var1}) comes from large $x$ which justifies the diffusion approximation employed
here. The first integral thus scales as $\sqrt{\tau_a}$ . Since $\tau_a \sim 1/\varepsilon_a$ we conclude that as $\varepsilon_a\downarrow 0$,
the $C_1$ contribution to the variance of LDOS vanishes. All diagrams with short-range spatial correlations in both source and receiver
positions have the same fate, in particular the diagram $C_1'$ In Fig.\ \ref{ldos}(b), which we discuss below.

We now turn to $C_2$, the first non-Gaussian contribution to the intensity correlation \cite{feng}. This is caused by a single crossing at an
arbitrary point $\mathbf{s}$ in the medium [see the second and the third diagrams in the left column of Fig.\ \ref{ldos}(b)], and is described
mathematically by the `Hikami box' vertex, with a scalar constant $\mathcal{H}$ that needs not be specified here. Two very similar contributions
exist that differ only in selection rules \cite{azi}. The first is short-range for $\mathbf{x}\neq \mathbf{x}'$, and equals
\begin{eqnarray}\label{C2a}
&&\langle I(\mathbf{r},\mathbf{x}) I(\mathbf{r},\mathbf{x'})
    \rangle_{C_{2a}} =
\mathcal{H}\pi^2
    \langle\rho(\mathbf{r})\rangle^2 \langle \mathrm{Im} \, G (\mathbf{x},\mathbf{x}')\rangle^2
\nonumber \\
&&\hspace*{5mm} \times \int d\mathbf{s}\,
    L^2(\mathbf{r},\mathbf{s}) (\nabla_1 \cdot \nabla_2) L(\mathbf{r}_1=\mathbf{s},\mathbf{x})
L(\mathbf{r}_2=\mathbf{s},\mathbf{x})
\nonumber \\
\end{eqnarray}
According to Eq.~(\ref{var}) we need the double integral $\varepsilon_a^2 \int d\mathbf{x} \int d\mathbf{x}'$ of this object and let
$\varepsilon_a $ tend to zero. One integral converges again rapidly after one extinction length and is finite without absorption. We shall write
$\int d\mathbf{x}' \langle\mathrm{Im}\, G(\mathbf{x},\mathbf{x}')\rangle^2 = V_0$ and rearrange expression (\ref{C2a}) to
\begin{eqnarray}\label{Var2a}
\mathrm{Var}_{2a}[\rho(\mathbf{r})] &=& \frac{\omega^4}{c^4} \mathcal{H} V_0   \langle
    \rho(\mathbf{r})\rangle^2 
\nonumber \\
&\times& \lim
    _{\varepsilon_a\downarrow 0}  \varepsilon_a^2  \int d\mathbf{x}
 \int d\mathbf{s}
L^2(\mathbf{r},\mathbf{s}) \, |\nabla
L(\mathbf{s},\mathbf{x})|^2
\nonumber \\
&\sim& V_0 \lim
    _{\varepsilon_a\downarrow 0}  \varepsilon_a^2  \int d\mathbf{x} \, |\nabla L(\mathbf{x})|^2
    \int d\mathbf{s}\,
L^2(\mathbf{s})
\nonumber \\
\end{eqnarray}
In the second step we conveniently made use of translational symmetry of the infinite medium. We see that $|\nabla L(\mathbf{x})| \sim 1/x^2 $
for large $x$, making the integral converge without the need of absorption. Its divergence for $x<\ell$ is an artifact of the diffusion
approximation which is not valid at small length scales, and which we shall ignore. Hence the volume integral over $\mathbf{x}$ is just finite,
without absorption. The integral over the position $\mathbf{s}$ of the Hikami box scales as $\sqrt{\tau_a}$. As $\varepsilon_a\downarrow 0 $ we
conclude that the contribution of the first $C_2$ term to the LDOS, $\mathrm{Var}_{2a}[\rho(\mathbf{r})]$, vanishes.

The second contribution from $C_2$ is long-range as a function of
$\mathbf{x} - \mathbf{x}'$ \cite{gabriel}. Its
expression reads
\begin{eqnarray}\label{C2b}
&&\langle I(\mathbf{r},\mathbf{x})
    I (\mathbf{r},\mathbf{x}') \rangle_{C_{2b}}
= \mathcal{H}\pi^4
    \langle\rho(\mathbf{r})\rangle^2 \langle\rho (\mathbf{x})\rangle^2
\nonumber \\
&&\hspace*{5mm} \times \int d\mathbf{s}\,
    L^2 (\mathbf{r},\mathbf{s}) (\nabla_1 \cdot \nabla_2)
L(\mathbf{r}_1=\mathbf{s},\mathbf{x})L(\mathbf{r}_2=\mathbf{s},\mathbf{x}')
\nonumber \\
\end{eqnarray}
and a little rearranging shows that its contribution to the
variance of LDOS is
\begin{equation}\label{Var2b}
 \mathrm{Var}_{2b}[\rho(\mathbf{r})] \sim
    \lim
    _{\varepsilon_a\downarrow 0}  \varepsilon_a^2  \left| \int d\mathbf{x} \, \nabla L(\mathbf{x}) \right|^2
    \int d\mathbf{s}\,
L^2(\mathbf{s})
\end{equation}
We note that $\int_V d\mathbf{x} \nabla L(\mathbf{x}) = \int_{A(V)} d\mathbf{A} \, L(\mathbf{x})$, where $A(V)$ is the surface enclosing the
volume $V$. The surface integral vanishes for any closed surface since $L(\mathbf{x})$ does not depend on the direction of $\mathbf{x}$. Thus,
$\mathrm{Var}_{2b}[\rho(\mathbf{r})]=0$.

The contribution of $C_3$-correlation, the origin of universal conductance fluctuations, can be handled similarly. $C_3$ contains two Hikami
boxes [$C_3$ in Fig.~\ref{ldos}(b)], but that is a technical complication, and in just the same way as for $C_2$ it can be shown to vanish as
$\varepsilon_a\downarrow 0$. The diagram $C_1'$ in Fig.~\ref{ldos}(b) looks very much like $C_3$ but has actually short spatial range in both
source and receiver positions. As a result it belongs to the class $C_1$, and its contribution to the variance of LDOS vanishes for the same
reason as seen in Eq.~(\ref{Var1}).

Finally, the $C_0$ correlation is  given by [see the lower right diagram in Fig.\ \ref{ldos}(b)]
\begin{eqnarray}
\label{C0}
\langle I(\mathbf{r},\mathbf{x})
I(\mathbf{r},\mathbf{x}') \rangle_{C_0}  = \langle I(\mathbf{r},\mathbf{x}) \rangle
\langle I(\mathbf{r},\mathbf{x}') \rangle \times C_0
\end{eqnarray}

\vspace*{0.1mm}
\noindent
with $C_0 $ a dimensionless scalar  depending on the nature of the scatterers. For weak white-noise, uncorrelated disorder $C_0 = \pi/k \ell$
\cite{shapiro,sergey}. The essential property of $C_0$ that is important here is its \emph{infinite} spatial range caused by the scattering of
waves going to arbitrarily distant $\mathbf{x}$ and $\mathbf{x}'$ on a common scatterer in the vicinity of the source at $\mathbf{r}$. Inserting
Eq.~(\ref{C0}) into the expression for the LDOS variance~(\ref{var}), and making use of Eq.~(\ref{twopar}) we obtain
\begin{eqnarray}\label{Var0}
    \mathrm{Var}_0[\rho(\mathbf{r})]&=& {\omega^2} C_0\pi^2\langle
    \rho(\mathbf{r})\rangle^2
\nonumber \\
&\times& \lim
    _{\varepsilon_a\downarrow 0}  \varepsilon_a^2  \left(  \int d\mathbf{x}  \, \langle \rho(\mathbf{x})\rangle
      \, L(\mathbf{r},\mathbf{x}) \right)^2\nonumber \\
&=&C_0\pi^2\langle
    \rho(\mathbf{r})\rangle^2  \langle \rho(\mathbf{x})\rangle^2\lim_{\varepsilon_a\downarrow 0}  \varepsilon_a^2
{\omega^2}K^2\tau_a^2
    \nonumber \\&=& C_0\, \langle
    \rho(\mathbf{r})\rangle^2
\end{eqnarray}
Hence $C_0$ provides the only surviving contribution to $\mathrm{Var}[\rho(\mathbf{r})]$.

\end{document}